\documentclass[a4paper,10pt]{article}
\usepackage[]{graphicx}

\def\si {\mathop{\mathrm{si}}\nolimits}
\def\ci {\mathop{\mathrm{ci}}\nolimits}

\def\Re {\mathop{\mathrm{Re}}\nolimits}
\def\Im {\mathop{\mathrm{Im}}\nolimits}

\def\tr {\mathop{\mathrm{Tr}}\nolimits}
\def\th {\tanh}

\def\PISQTW {\frac{\pi^{2}}{12}}
\def\QSQTW {\frac{q_{\bot}^{2}}{12}}

\title{The qubit states decoherence in antiferromagnet-based nuclear spin model of quantum register}
\author{Alexander A.Kokin$^{1}$\footnote{E-mail: aakokin@mail.ru}~ and Vladimir A.Kokin$^{2}$}

\date{}

\begin{document}

\maketitle
\thanks{
$^{1}$Institute of Physics and Technology of RAS, 34, Nakhimovskii pr., 117218 Moscow, Russia;

$^{2}$Institute of Radioengineering and Electronics of RAS, 11, Mokhovaya str., 103907, Moscow, Russia
}

\begin{abstract} 
This study deals with the further development of nuclear spin model of scalable quantum register, 
which presents the one-dimensional chain of the magnetic atoms with nuclear spins 1/2, 
substituting the basic atoms in the plate of nuclear spin-free easy-axis 3D antiferromagnet. 
The decoherence rates of one qubit state and entanglement state of two removed qubits and 
longitudinal relaxation rates are caused by the interaction of nuclear spins-qubits with virtual 
spin waves in antiferromagnet ground state were calculated. 
It was considered also one qubit adiabatic decoherence, is caused by the interaction of nuclear spin 
of quantum register with nuclear spins of randomly distributed isotopes, 
substituting the basic nuclear spin-free isotopes of antiferromagnet. 
We have considered finally encoded DFS (Decoherence-Free Subspaces) 
logical qubits are constructed on clusters of the four-physical qubits, 
given by the two states with zero total angular momentum.
~


\end{abstract}

PACs: 75.10.Pq, 75.50.Ee, 76.60.-k, 82.56.-b.

\section{Introduction}

In early papers [1-4], we have considered a model of NMR quantum register,
which is based on the nuclear spin-free easy-axis 3D antiferromagnet at low temperature
in homogeneous field. It was shown that the range of indirect coupling can
ran up to a great value close to critical point of spin-flop quantum phase transition
in antiferromagnet. We have extended the previous model into the case of inhomogeneous
external magnetic field.

It was proposed to use the natural antiferromagnetic crystals
with easy-axis anisotropy as an antiferromagnet thin plate (or film).
As examples, they may be crystals CeC$_{2}$ with tetragonal and FeCO$_{3}$ (siderite)
with trigonal symmetry. The basic isotopes of these crystals $^{12}\mathrm{C}$, $^{56}\mathrm{fe}$ (91.7\%),
$^{16}\mathrm{O}$, $^{140,142}\mathrm{Ce}$ (99.6\%) have no nuclear spins 
(in brackets percent isotopic abundance is given). To form the one-dimension nuclear spin chains the isotopic substitution atoms,
such as $^{12}\mathrm{C}$ in corresponding crystal lattice sites, for isotopes $^{13}\mathrm{C}$
with nuclear spins 1/2, are proposed. One would expect that period of such solid
state NMR quantum registers may be much more than periods of crystal lattice.

The simple antiferromagnet model to be studied here consists of two
incorporated to each other tetragonal magnetic sublattices
\textbf{A} and \textbf{B} with $N = N_{\bot} N_{z} $
sites in each sublattice, where $N_{\bot} = N_{x} N_{y} \gg 1$ are the sites numbers
in plane of plate ($x$,$y$-axes) and $N_{z} > 1$ is the sites numbers in $z$ direction.
The atom sites of sublattices are numbered respectively by numbers $j$ and $i$.
Each sublattice constant in the plane of the plate is $a_{\bot}$ and along symmetry axis is $a_{z}$.

\begin{figure}
 \begin{center}
 \begin{tabular}{c}
 \includegraphics{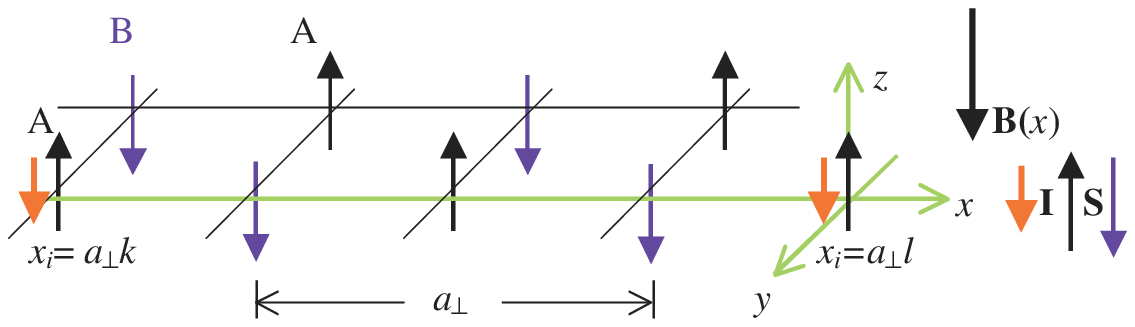}
 \end{tabular}
 \end{center}
 \caption[1]
%>>>> use \label inside caption to get Fig. number with \ref{}    Fig.~\ref{fig1}
 {\label{fig:1}
The scheme of antiferromagnet based nuclear spin quantum register in external field
lower than critical field for spin-flop phase transition $B\left({x}\right) \ll B_{C}$.
The oriented (that is nonprecessing) arrows represent here the ground states
of corresponding individual Bloch vectors.
The nuclear spins \textbf{I}$_{A}$ (shot red arrows) are contained here only
in atoms of sublattice~\textbf{A}.
 }
 \end{figure}

The external magnetic field $B\left({x}\right)$ in considered model
(Fig.~\ref{fig:1}) is directed parallel to z-axis and to the antiferromagnet easy axis.
It is assumed here, that the field gradient is of the order
$dB_{z} \left({x}\right)/dx = G \sim 0.1\,\mathrm{T}/\mu\mathrm{m}$
($x-$axis in plane of the plate along nuclear spin chain).
This value of the gradient corresponds to the difference of resonance
frequencies of the order of 100 kHz for two nuclear spins,
being separated by $100\,a_{\bot}$ ($a_{\bot} \sim 1\,\mathrm{nm}$).

%-------------
%-------------

The qubits number in quantum register will be limited by planar structure dimensions. For example, the structure with linear dimension of the order of $10\,\mu \mathrm{m}$ will have 100 qubit register with period $L=100a_{\bot}$.

The starting spin Hamiltonian of 3D easy-axis antiferromagnet with interaction only between neighbouring atoms, which belong to the distinct sublattice, is represented for our model~as 
\begin{equation}
\label{eq1}
H_{S} = \gamma_{S} \hbar \left({\sum\limits_{i}^{N} {B\left({x_{i}}\right)\;S_{Az} \left({r_{i}}\right)} + \sum\limits_{j}^{N} {B\left({x_{j}}\right)\;S_{Bz}} \left({r_{j}}\right)\;}\right) + 
\end{equation}
\nopagebreak $$ % continuation of eq1
 + 2 \gamma_{S} \hbar /Z\;\sum\limits_{i}^{N} {\;\sum\limits_{\delta} ^{Z} {\{ \;B_{E} \;S_{A} \left({r_{i}}\right)\;S_{B} \left({r_{i} + r_{\delta }}\right)\; + \;B_{A} \;S_{Az} \left({r_{i}}\right)\;S_{Bz} \left({r_{i} + r_{\delta} }\right)\} }}, 
$$ 
\noindent
where $S_{A} \left({r_{i}}\right)$ 
and 
$S_{B} \left({r_{i} + r_{\delta} }\right)$ 
are electron spin operators (\textit{S} = 1/2) for neighbouring sites of sublattices \textbf{A} and \textbf{B}, 
$B\left({x_{i}}\right) = B + Gx_{i}$, $B$ is the field value at the origin of the coordinates $x_{i}$,
 $Z = 6$ is the number of neighbouring atoms for tetragonal sublattice, 
$\gamma_{S} = 175.88\, \mathrm{rad\,GHz/T}$ $\left({\gamma_{S} /2\pi = 28\,\mathrm{GHz/T}}\right)$ 
is electron spin gyromagnetic ratio, $\hbar = 1.054 \cdot 10^{-34}\,\mathrm{J s/rad}$.

The direct product of spin operators in Eq.(\ref{eq1}) written in matrix representation 
is common designated by symbol $\otimes$. 
In the following this symbol will be for brevity omitted.

In Eq.(\ref{eq1}) the parameters
$B_{E}$ $\sim$ $\left({10 - 100}\right)\mathrm{T}$, 
$B_{A}$ $\sim$ $\left({10^{ - 2} - 1}\right)\mathrm{T} > 0$, 
$B_{C} = \sqrt {2B_{E} B_{A} + B_{A}^{2}}$ 
are exchange field, anisotropy and critical spin-flop field for easy-axis antiferromagnet (particularly, 
for $\mathrm{FeCO}_{3}$: $B_{E} = 35\;\mathrm{T},\;\;B_{A} = 3.3\,\mathrm{T}$ $B_{C} \approx 15.5\,\mathrm{T}$).

It was shown in Ref.[1-4], that indirect interaction of two nuclear spins in the model 
of antiferromagnet-based nuclear spin quantum register with inhomogeneous magnetic field 
essentially grows and qualitatively changes its character, 
if the value of local field in the mid point of two considered nuclear spin 
is close to the critical field for quantum phase transition of spin-flop type 
in bulk easy-axis antiferromagnet. 
The corresponding coordinates of nuclear spins were named by us as turning points.

In present paper, we have presented the further investigations and development of this model. 
With the more refined results of the model analysis it was investigated here in details 
the one-qubit and two-qubit nonadiabatic decoherence and longitudinal relaxation rates are caused 
by the interaction of nuclear spins with virtual spin waves in antiferromagnet ground state. 
It was considered also one qubit adiabatic decoherence, is caused by the interaction with 
nuclear spins of randomly distributed isotopes, substituting the nuclear spin-free isotopes 
of basic antiferromagnet. We have considered finally, as an example, 
encoded DFS (Decoherence-Free Subspaces)-states of logical qubits are constructed 
on clusters of the four-physical qubits, given by the two states with zero total angular momentum.

The paper is organized as follows. 
In Section 2 we have considered some general expressions of one qubit decoherence 
and relaxation processes are caused by interaction of nuclear spin with spin waves 
in easy axis antiferromagnet plate with inhomogeneous magnetic field. 
In Section 3 it was considered the nonadiabatic one qubit decoherence and longitudinal relaxation. 
In Section 4 it was considered the decoherence of two qubit entangled quantum states. 
In Section 5 we have considered the adiabatic decoherence is caused by interaction with 
nuclear spins of random distributed isotopes, substituting the spin-free atoms of basic antiferromagnet. 
In Section 6 we have discussed an encoded DFS (Decoherence-Free Subspaces)-states of logical qubits 
on clusters of the four-physical qubits. 
In Conclusion we discuss the some prospects of considered quantum register model.

\section{Some general expressions for one qubit decoherence and longitudinal relaxation 
in antiferromagnet-based quantum register}

The antiferromagnet electron spin system of considered model plays the role of an environment 
for nuclear spin quantum register, whose interaction with spin wave leads on the one hand, 
to indirect coupling between nuclear spins and on the other hand to decoherence and relaxation 
processes of their states.

Let us consider the decoherence and relaxation processes of quantum state for single nuclear 
spin placed at position $k$ on axes $x$ of sublattice \textbf{A} 
which is caused by its interaction with virtual magnon excitation in antiferromagnet. 
Such processes are described by transverse and longitudinal relative to external field components of Bloch vector (Ref.[5]) 
$$ % beginning of eq2
 \,\,P^{-} \left({k,\tau}\right) =
 \left({P^{+} \left({k,\tau}\right)}\right)^{ *} =
 P_{x} \left({k,\tau}\right) - iP_{y} \left({k,\tau}\right) =
 2 \tr\left({I^{-} \left({k}\right)\rho \left({k,\tau}\right)}\right) =
 2 \tr_{I} \left({I^{-} \left({k}\right)\rho _{I} \left({k,\tau}\right)}\right)
$$ 
\nopagebreak \begin{equation}
\label{eq2}
 \,P_{z} \left({k,\tau}\right) =
 2 \tr_{I} \left({I_{z} \left({k}\right)\rho _{I} \left({k,\tau}\right)}\right),\,
 \tr_{I} I_{z}^{2} = 1/2, \tr_{I} I^{+} I^{-} = 1,
\end{equation}
\noindent
where non-steady reduced to one nuclear spin density matrix in antiferromagnet is represented by
\begin{equation}
\label{eq3}
\rho_{I} \left({k,\tau}\right) =
1/2\left[ {1 + P_{z}\left({k,\tau}\right) I_{z}\left({k,\tau}\right) +
P^{-}\left({k,\tau}\right)I^{+}\left({k,\tau}\right) +
P^{+}\left({k,\tau}\right)I^{-}\left({k,\tau}\right)
}\right]
\end{equation}

Let us pass next in Eq. (\ref{eq1}) to dimensionless designations:
\begin{equation}
\label{eq4}
h_{S} = H_{S} /\hbar \omega_{E} ,
\,\gamma_{S} B_{E} = \omega_{E} ,
\,0 < B_{A} /B_{E} = b_{A} < 1,
0 < B/B_{E} = b,\,\,\,\,G a_{\bot} /B_{E} = g \sim 10^{ - 5}\,.
\end{equation}

The interaction of \textit{k}-th nuclear spin with external magnetic field and magnon excitations will be described here by dimensionless Hamiltonian with hyperfine interaction of the form (Ref.[4])

$$ % beginning of eq5
 h = H/\hbar \omega_{E} = h_{S} - \left({\omega_{I} \left({k}\right) - aS_{z} \left({k}\right)}\right)I_{z} \left({k}\right) + a/2\left({I^{+} \left({k}\right)S^{-} \left({k}\right) + I^{-} \left({k}\right)S^{+} \left({k}\right)}\right) = 
$$ \nopagebreak \begin{equation}
\label{eq5}
 = h_{S} - \left({\omega_{I} \left({k}\right) - a\left({1/2 - \psi}\right)I_{z} \left({k}\right) + \Delta h_{IS} \left({k}\right)
 }\right)
 \,,
\end{equation}
\noindent
where 
$I_{z}$, $I^{ \pm} = I_{x} \pm iI_{y}$, $S_{z}$, $S^{ \pm} = S_{x} \pm iS_{y}$
are nuclear and electron spin operators,
$\omega_{I} \left({k}\right) = \left({\gamma_{I} /\gamma_{S}}\right)b_{k}$
is resonance nuclear frequency in local field
$b_{k} \equiv b + gk$,
$\gamma_{I} /\gamma_{S} \sim 10^{ - 3}$,
$a = A/\omega_{E} \sim 10^{ - 3} - 10^{ - 4}$
is isotropic dimensionless constant of hyperfine interaction and
\begin{equation}
\label{eq6}
\langle 0|1/2 - S_{z} \left({k}\right)|0\rangle \approx \langle 0|S^{ + }\left({k}\right)S^{-} \left({k}\right)|0\rangle = \psi \left({k}\right) \ll 1.
\end{equation}
\noindent
is known as ``spin contraction''.

The perturbation Hamiltonian, corresponding to the relaxation and decoherence processes in isotopic pure antiferromagnet, has the form
\begin{equation}
\label{eq7}
\Delta h_{IS} \left({k}\right) = \Delta h_{IS}^{\left({1}\right)} \left({k}\right) + \Delta h_{IS}^{\left({2}\right)} \left({k}\right),
\end{equation}
\noindent
where
\begin{equation}
\label{eq8}
\Delta h_{IS}^{\left({1}\right)} \left({k}\right) = a/2\left({I^{ + }\left({k}\right)S^{-} \left({k}\right) + I^{-} \left({k}\right)S^{+} \left({k}\right)}\right)
\end{equation}

The second term in Eq.(\ref{eq7})
\begin{equation}
\label{eq9}
\Delta h_{IS}^{\left({2}\right)} \left({k}\right) = aI_{z} \,\left({k}\right)\left({S_{z} \left({k}\right) - \langle 0|S_{z} \left({k}\right)|0\rangle}\right) \approx - aI_{z} \left({k}\right)\left({\,S^{-} \left({k}\right)S^{+} \left({k}\right) - \psi}\right)
\end{equation}
\noindent
describes two-magnon interaction, which is similar to two-phonon interaction (Ref.[5],\S 3.4). 
This mechanism causes in particular the modulation of nuclear spin resonance frequency without 
changing its state (adiabatic decoherence). It leads to the temperature depending decoherence rate, 
which is negligible small at 
$\left({b_{C} - b \gg \frac{{k_{B} T}}{{\hbar \gamma_{S} B_{C}} }}\right)$, 
when the system is close to ground state. 
Therefore, we will neglect next the contribution of terms 
$\Delta h_{IS}^{\left({2}\right)} \left({k}\right)$ 
and will use as the perturbation Hamiltonian only the expression 
$\Delta h_{IS} \left({k}\right) = \Delta h_{IS}^{\left({1}\right)} \left({k}\right)$. 
In this case, relaxation of transverse component of Bloch vector is accompanied 
by nuclear spin flopping (nonadiabatic decoherence). 
At the same time, the relaxation of longitudinal component of Bloch vector also occurs. 
Thus, these two processes may be considered here as one unified process of nuclear quantum state damping.

We assume now that interaction of nuclear spin, which is initially at coherent state (with nonzero nondiagonal elements of density matrix $\rho _{I} \left({k,0}\right)$), and antiferromagnet in ground state, is turning on at the initial moment $\tau = 0$, when nonperturbed density matrix is represented as direct product 
$\rho \left({k,0}\right) = \rho _{I} \left({k,0}\right)\rho _{S} \left({0}\right) = \rho _{I} \left({k,0}\right)|0\rangle \langle 0|$.

Let us go next to interaction representation for density matrix relatively to Hamiltonian 
$h_{0} \left({k}\right) = h_{S} - \left({\omega_{I} \left({k}\right) - a/2}\right),$ 
$I_{z} \left({k}\right)$:
\begin{equation}
\label{eq10}
\rho _{in} \left({k,\tau}\right) = \exp\left({ih_{0} \left({k}\right)\tau}\right)\rho \left({k,\tau}\right)\exp\left({ - ih_{0} \left({k}\right)\tau}\right)
\end{equation}
\noindent
and to the equation for density matrix of nucleus-electron system
\begin{equation}
\label{eq11}
i\partial \rho _{in} \left({k,\tau}\right)/\partial \tau = \left[ {\Delta h_{IS} \left({k,\tau}\right),\,\rho _{in} \left({k,\tau}\right)}\right],
\end{equation}
\noindent
where
\begin{equation}
\label{eq12}
 \Delta \,h_{IS} \left({k,\tau}\right) =
 \exp\left[ {i h_{0} \left({k}\right)\tau}\right]\,\Delta h_{IS} \left({k}\right)\exp\left[ { - i h_{0} \left({k}\right)\tau}\right] = 
\end{equation}
\nopagebreak $$ % continuation of eq1
 = a/2\,\,\exp\left({ - i\left({\omega_{I} \left({k}\right) - a/2}\right)\tau}\right)\,I^{+} \left({k}\right)S^{-} \left({k,\tau}\right) + H.c.
$$

It is follows from Eq.(\ref{eq11}) in the second order of perturbation theory
\begin{equation}
\label{eq13}
i\partial \rho _{in} \left({k,\tau}\right)/\partial \tau \approx
\left[ {\Delta h_{IS} \left({k,\tau}\right),\rho \left({k,0}\right)}\right]
- i\int\limits_{0}^{\tau} {\left[ {\Delta h_{IS} \left({\tau ,k}\right),\left[ { \Delta h_{IS} \left({k,{\tau}'}\right),\rho \left({k,0}\right)}\right]}\right]}d{\tau} '.
\end{equation}

Let us write the derivative of expression (\ref{eq2}) 
with respect to time, by using Eq.(\ref{eq13}) and perform then the cyclic permutation under tracing. 
Finally, accounting the relation \linebreak
$\tr\left({\left[{I^{-}\left({k}\right),\Delta h_{IS} \left({k,\tau}\right)}\right]\rho_{S}\left({0}\right)}\right) = 0$, 
we will find
\begin{equation}
\label{eq14}
 \partial \left[ {P^{-} \left({k,\tau}\right)\exp\left({i\left({\omega_{I} \left({k}\right) - a\left({1/2}\right)}\right)\tau}\right)}\right]/\partial \tau = 2 \tr\{ I^{-} \left({k}\right)\partial \rho _{in} \left({k,\tau}\right)/\partial \tau \} \approx
\end{equation}
\nopagebreak $$ % continuation of eq14
 \approx - 2 \tr\{ \int\limits_{0}^{\tau} {\left[ {\left[ {I^{-} \left({k}\right),\Delta h_{IS} \left({k,\tau}\right)}\right],\,\,\Delta h_{IS} \left({k,{\tau} '}\right)}\right]\rho \left({k,0}\right)\}} d{\tau} '.
$$

By defining the transverse Bloch vector component in the form
\begin{equation}
\label{eq15}
P^{-} \left({k,\tau}\right) = P^{-} \left({k,0}\right)\exp\left[ { - i\left({\omega_{I} \left({k}\right) - a/2}\right)\tau - \gamma_{\bot} \left({k,\tau}\right)}\right],
\end{equation}
\noindent
we will have:
\begin{equation}
\label{eq16}
d\left[ {P^{-} \left({k,\tau}\right)\exp\left({i\left({\omega_{I} \left({k}\right) - a/2}\right)\tau}\right)}\right]/d\tau =
\end{equation}
\nopagebreak $$ % continuation of eq16
= - d\gamma_{\bot} \left({k,\tau}\right)/d\tau \, \cdot P^{-} \left({k,\tau}\right)\exp\left({i\left({\omega_{I} \left({k}\right) - a/2}\right)\tau}\right), 
$$
\noindent
where $\Re\gamma_{\bot} \left({\tau}\right)$ is decoherence decrement and 
$\Im\gamma_{\bot} \left({\tau}\right)$ is phase shift.

We will represent the nuclear density matrix in the right part of Eq.(\ref{eq14}) by expression
\begin{equation}
\label{eq17}
\rho _{I} \left({k,0}\right) \approx 1/2\{ 1 + 2P_{z} \left({k,0}\right)I_{z} \left({k}\right) + P^{-} \left({k,0}\right)I^{+} \left({k}\right) + P^{+} \left({k,0}\right)I^{-} \left({k}\right)\} |0\rangle \langle 0|.
\end{equation}

Using in Eq.(\ref{eq14}) the perturbation Hamiltonian (\ref{eq8}), ignoring the factors 
\linebreak $\exp\left({ \pm i\left({\omega_{I} \left({k}\right) - a/2}\right)\tau }\right)$ 
and taking in the context of second order of perturbation theory in the left-hand side of Eq.(\ref{eq16}) $P^{-} \left({k,\tau}\right) \approx P^{-} \left({k,0}\right)$, for the decoherence rate we will obtain
\begin{equation}
\label{eq18}
d\Re\gamma_{\bot} \left({k,\tau}\right)/d\tau = \Re \tr_{I} \int\limits_{0}^{\tau} {\langle 0|\left[ {\left[ {I^{-} \left({k}\right),\Delta h_{IS} \left({k,\tau}\right)}\right],\,\,\Delta h_{IS} \left({k,{\tau} '}\right)}\right]\,} I^{ + }\left({k}\right)|0\rangle d{\tau} ' =
\end{equation}
\nopagebreak $$ % continuation of eq18
= \frac{{a^{2}}}{{4}}2\Re\int\limits_{0}^{\tau} {\langle 0|S^{-} \left({k,\tau}\right)S^{-} \left({k,{\tau} '}\right) + S^{+} \left({k,{\tau }'}\right)S^{-} \left({k,\tau}\right)|0\rangle d{\tau} '} . 
$$

Let us consider next the relaxation longitudinal component of Bloch vector. We will write:
\begin{equation}
\label{eq19}
\partial P_{z} \left({k,\tau}\right)/\partial \tau = 2 \tr I_{kz} \partial \rho _{in} \left({k,\tau}\right)/\partial \tau \approx 
\end{equation}
\nopagebreak $$ % continuation of eq19
\approx - \Re 2 \tr \int\limits_{0}^{\tau} {\left[ {\left[ {I_{kz} ,\Delta h_{IS} \left({k,\tau}\right)}\right],\,\,\Delta h_{IS} \left({k,{\tau} '}\right)}\right]\rho _{I} \left({k,0}\right)} |0\rangle \langle 0|d{\tau} '.
$$ 

Taking $P_{z} \left({k,\tau}\right) = P_{z} \left({k,0}\right)\exp\left({ - \gamma_{\parallel} \left({k,\tau}\right)}\right)$, 
in the second order of the perturbation theory we will obtain:
\begin{equation}
\label{eq20}
dP_{z} \left({k,\tau}\right)/d\tau \approx - d\gamma_{\parallel} \left({k,\tau }\right)/d\tau \, \cdot P_{z} \left({k,0}\right),
\end{equation}
\noindent
where the longitudinal relaxation rate is
\begin{equation}
\label{eq21}
d\gamma_{\parallel} \left({k,\tau}\right)/d\tau = \left({d\gamma_{\parallel} \left({k,\tau }\right)/d\tau}\right)^{ *} =
\end{equation}
\nopagebreak $$ % continuation of eq21
= 2 \Re \tr_{I} \int\limits_{0}^{\tau} {\langle 0|\left[ {\left[ {I_{z} \left({k}\right),\Delta h_{IS} \left({k,\tau}\right)}\right],\,\,\Delta h_{IS} \left({k,{\tau} '}\right)}\right]\,I_{z} \left({k}\right)|0\rangle} d{\tau} '. 
$$

As a result of Eq.(\ref{eq21}) we will have
\begin{equation}
\label{eq22}
d\gamma_{\parallel} \left({k,\tau}\right)/d\tau \approx 
\end{equation}
\nopagebreak $$ % continuation of eq22
\approx \frac{{a^{2}}}{{4}}2\Re\int\limits_{0}^{\tau} {\langle 0|S^{-} \left({k,\tau}\right)S^{+} \left({k,{\tau} '}\right) + S^{+} \left({k,{\tau }'}\right)S^{-} \left({k,\tau}\right)|0\rangle d{\tau} '} \, = \Re d\gamma_{\bot} \left({k,\tau}\right)/d\tau ,
$$
\noindent
whence it follows that for the considered mechanisms the rate of 
relaxation of longitudinal component is equal to the rate of relaxation 
of transverse component (decoherence rate).

\section{Nonadiabatic decoherence and longitudinal relaxation rates of one qubit quantum states}

For the calculation of nonadiabatic one qubit decoherence and longitudinal 
relaxation rates we used next the results of antiferromagnet Hamiltonian 
diagonalization, obtained in Ref.[4]. The transverse components of electron spin operators 
will take the form
\begin{equation}
\label{eq23}
S^{-} \left({k,\tau}\right) = \left({S^{+} \left({k,\tau}\right)}\right)^{+} = \\
\end{equation}
\nopagebreak $$ % continuation of eq23
= \frac{{1}}{{\left({2\pi}\right)}}\int {\left[ {u^{*} \left({q_{\bot},E}\right)\exp\left({iE_{-} \tau}\right)\xi ^{+} \left({q_{y},E_{-} }\right) + v\left({q_{\bot},E}\right)\exp\left({ - iE_{+} \tau }\right)\xi \left({q_{y},E_{+} }\right)}\right]\exp\left({iq_{x} k}\right)dEdq_{\bot} } , 
$$
\noindent
where $\xi ^{+} \left({q_{y} ,E_{-} }\right)$, $\xi \left({q_{y} ,E_{+} }\right)$
are operators of creation and annihilation of spin magnons for two branches of magnon states, 
which propagate along direction \textit{x}-axis with dimensionless energies 
$E_{ \pm} = E \pm b$, 
$E$ is a continuous energy parameter, and wave vector component 
$q_{y}$ in the range of $q_{y}$ to $q_{y} + dq_{y}$. 
The transformation coefficients have here the following asymptotic form (Ref.[4])
\begin{equation}
\label{eq24}
 u^{*}\left({q_{\bot},E}\right) = \frac{{1}}{{\sqrt {4\pi g}} }\sqrt {\frac{{\sqrt {1 + b_{C}^{2} } }}{{E}} + 1} \cdot \exp\left[ {\frac{{i}}{{g}}\left({Eq_{x} - \int\limits_{0}^{q_{x}} {E\left({q_{\bot}}\right)\;dq_{x}} }\right)}\right] + O\left({g}\right), 
\end{equation}
\nopagebreak $$ % continuation of eq24
 v^{*}\left({q_{\bot},E}\right) = \frac{{1}}{{\sqrt {4\pi g}} }\sqrt {\frac{{\sqrt {1 + b_{C}^{2} } }}{{E}} - 1} \cdot \exp\left[ {\frac{{i}}{{g}}\left({Eq_{x} - \int\limits_{0}^{q_{x}} {E\left({q_{\bot}}\right)\;dq_{x}} }\right)}\right] + O\left({g}\right),
$$
\noindent
where $b_{C} = B_{C} /B_{A}$ is dimensionless critical field of phase transition and 
$E\left({q_{\bot} }\right) \approx \sqrt {b_{C}^{2} + \QSQTW}$.

Let us restrict next to low magnon excitation mode with energy $E_{-}$.
Further, let us insert week magnon damping $s$ 
($E_{-} \to E_{-} + is$, $E_{-} \gg s > 0$)
and make a set of rearrangements, after which we will obtain
\begin{equation}
\label{eq25}
\Re d\gamma_{\bot} \left({k,\tau}\right)/d\tau \approx 
\end{equation}
\nopagebreak $$ % continuation of eq25
\approx \frac{{a^{2}}}{{\left({4\pi}\right)^{2}}}\Re 2\int {\int\limits_{0}^{\tau} {u^{*} \left({q_{\bot},E}\right)u\left({{q}'_{x},q_{y},E}\right)\exp\left({ - \left({iE_{-} - s}\right)\left({{\tau} ' - \tau}\right)}\right)\exp\left({i\left({q_{x}-{q}'_{x}}\right)k}\right)d{q}'_{x} dq_{\bot} dEd{\tau} '}}.
$$ 

For the estimation of magnon damping it may be used the line width of antiferromagnetic resonance 
$s \sim \gamma_{S} \Delta B/\omega_{E}$. 
The typical value of AFR line width is $\Delta B \sim 10^{-4}\;\mathrm{T}$ and $s \sim \Delta B/B \sim 10^{-5}$.

Upon integrating over ${\tau}'$, we have 
\begin{equation}
\label{eq26}
\Re d\gamma_{\bot} \left({k,\tau}\right)/d\tau \approx 
\end{equation}
\nopagebreak $$ % continuation of eq26
 \approx \frac{{a^{2}}}{{\left({4\pi}\right)^{2}}}\Re 2\int {u^{ *} \left({q_{\bot} ,E}\right)u\left({{q}'_{x} ,q_{y} ,E}\right)\exp\left({i\left({q_{x} - {q}'_{x}}\right)k}\right)\frac{{1 - \exp\left({\left({iE_{-} - s}\right)\tau}\right)}}{{ - iE_{-} + s}}dEdq_{\bot} d{q}'_{x}} . 
$$

By using the expression 
$E\left({q_{\bot} }\right) \approx \sqrt {b_{C}^{2} + \QSQTW}$, 
we write
\begin{equation}
\label{eq27}
dq_{\bot} = 2\pi q_{\bot} dq_{\bot} = 24\pi E\left({q_{\bot} }\right)dE\left({q_{\bot} }\right),\,\,\,\,\,0 < q_{\bot} \le \pi \end{equation}
\noindent
and take notations 
$\Delta b_{k} = b_{C} - b - gk$, $\,\xi = E\left({q_{\bot} }\right)- b_{C}$. 
Upon integrating over $\,\,E$ and ${q}'_{x} $ (Ref.[4]), we will transform Eq.(\ref{eq26}) for decoherence rate to the form
\begin{equation}
\label{eq28}
\Re d\gamma_{\bot} \left({k,\tau}\right)/d\tau = \frac{{3a^{2}}}{{2\pi} }R_{\bot} \left({\Delta b_{k} ,\tau}\right) =
\end{equation}
\nopagebreak $$ % continuation of eq26
= \frac{{3a^{2}}}{{2\pi} }\,\,\int\limits_{0}^{\sqrt {b_{C}^{2} + \PISQTW} - b_{C}} {\left({\sqrt {1 + b_{C}^{2}} + b_{C} + \xi}\right)\,\,Y\left({\xi , + \Delta b_{k} ,\tau}\right)\,d\xi} ,
$$ 
\noindent
where
\begin{equation}
\label{eq29}
Y\left({\xi + \Delta b_{k} ,\tau}\right) = \Re\frac{{1 - \exp\left[ {\left({i\left({\xi + \Delta b_{k}}\right) - s}\right)\tau}\right]}}{{ - \left({i\left({\xi + \Delta b_{k}}\right) - s}\right)}} = 
\end{equation}
\nopagebreak $$ % continuation of eq26
 = \frac{{\left({\xi + \Delta b_{k}}\right)\sin\left({\left({\xi + \Delta b_{k}}\right)\tau}\right)\exp\left({ - s\tau}\right)}}{{\left({\xi + \Delta b_{k}}\right)^{2} + s^{2}}} + s\frac{{1 - \cos\left({\left({\xi + \Delta b_{k}}\right)\tau}\right)\exp\left({ - s\tau}\right)}}{{\left({\xi + \Delta b_{k}}\right)^{2} + s^{2}}}
$$ 

For the case of $\Delta b_{k} \gg s > 0$, after omitting $s^{2}$ in dominator the Eq.(\ref{eq28}) take the following explicit form.

$$ % begining of eq30
R_{\bot} \left({\Delta b_{k},\tau}\right) \approx \{ \left({\sqrt {1 + b_{C}^{2}} + b_{C} - \Delta b_{k}}\right)\left[ {\si\left({\Delta b_{k} \tau}\right) - \si\left({\left({\sqrt {b_{C}^{2} + \PISQTW} - b_{C} + \Delta b_{k}}\right)\tau}\right)}\right] +
$$
\nopagebreak $$ % continuation of eq30
 + 1/\tau \left[ {\cos\left({\Delta b_{k} \tau}\right) - \cos\left({\left({\sqrt {b_{C}^{2} + \PISQTW} - b_{C} + \Delta b_{k}}\right)\tau}\right)}\right]\} \exp\left({-s\tau}\right) +
$$
\nopagebreak $$ % continuation of eq30
 + s \left({\sqrt{1 + b_{C}^{2}} + b_{C} - \Delta b_{k}}\right)\{ \frac{{\sqrt {b_{C}^{2} + \PISQTW} - b_{C}} }{{\Delta b_{k} \left({\sqrt {b_{C}^{2} + \PISQTW} - b_{C} + \Delta b_{k}}\right)}} - \left[ {\frac{{\cos\left({\Delta b_{k} \tau}\right)}}{{\Delta b_{k}} } - \frac{{\cos\left({\left({\sqrt{b_{C}^{2} + \PISQTW} - b_{C} + \Delta b_{k}}\right)\tau}\right)}}{{\sqrt{b_{C}^{2} + \PISQTW} - b_{C} + \Delta b_{k}} }
 }\right.+ 
$$
\nopagebreak \begin{equation}
\label{eq30}
 + \left.{
 \tau \left({\si\left({\Delta b_{k} \tau}\right) - \si\left({\left({\sqrt {b_{C}^{2} + \PISQTW} - b_{C} + \Delta b_{k}}\right)\tau}\right)}\right)}\right] \exp\left({-s\tau}\right)\} + 
\end{equation}
\nopagebreak $$ % continuation of eq30
+ s\,\{ \log\frac{{\left({\sqrt {b_{C}^{2} + \PISQTW} - b_{C} + \Delta b_{k}}\right)}}{{\Delta b_{k}} } + \,\left[ {\ci\left({\Delta b_{k} \tau}\right) - \ci\left({\left({\sqrt{b_{C}^{2} + \PISQTW} - b_{C} + \Delta b_{k}}\right)\tau}\right)}\right] \exp\left({-s\tau}\right)\} > 0, 
$$ 
\noindent
where 
$\si\left({x}\right) = -\int\limits_{x}^{\infty}{\frac{{\sin t}}{{t}}dt}$, 
$\ci\left({x}\right) = -\int\limits_{x}^{\infty}{\frac{{\cos t}}{{t}}dt}$ 
are sine-integral, and cosine-integral, and 
\linebreak
$\mathop {lim}\limits_{\tau \to 0} \left[ {\ci\left({a\tau}\right) - \ci\left({b\tau}\right)}\right] = \log\frac{{a}}{{b}}$.

\begin{figure}
 \begin{center}
 \begin{tabular}{c}
 \includegraphics{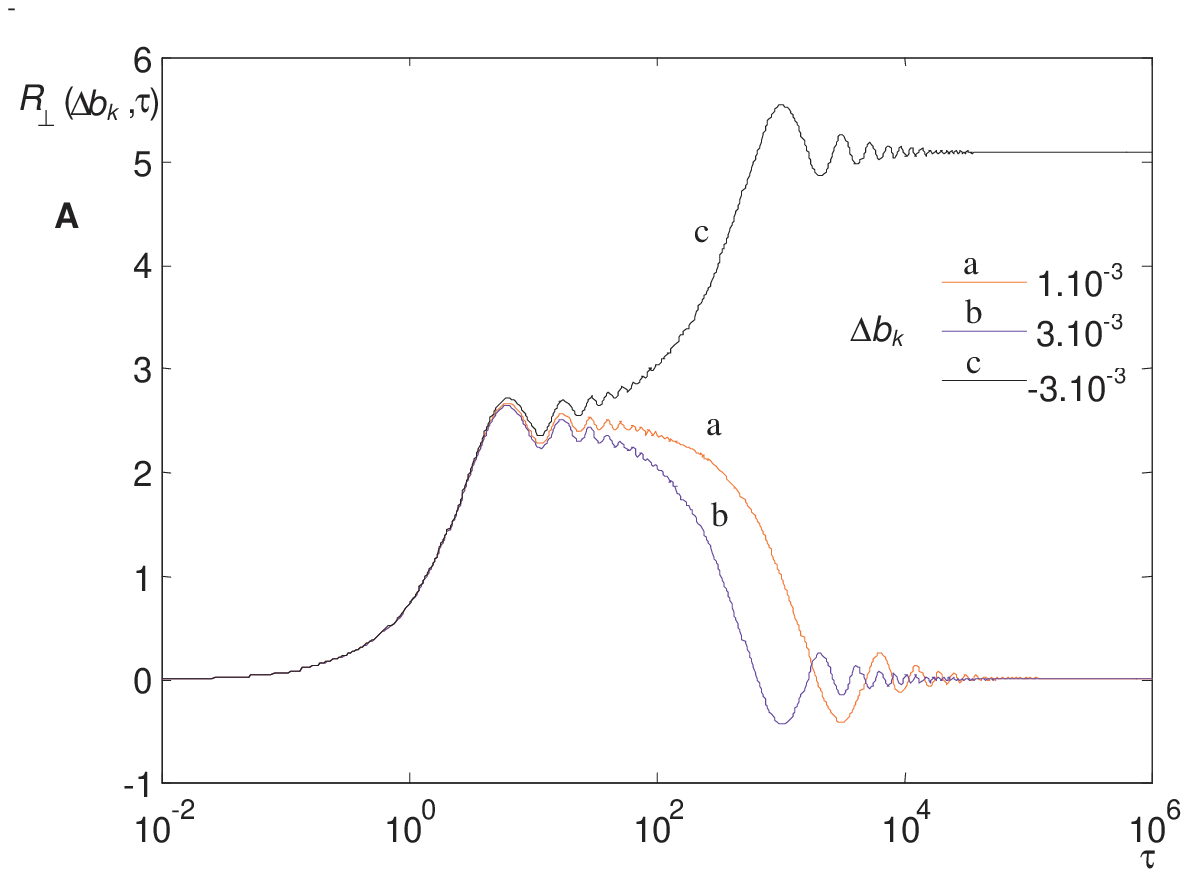}
\\
~~~~~~~~~~
 \includegraphics{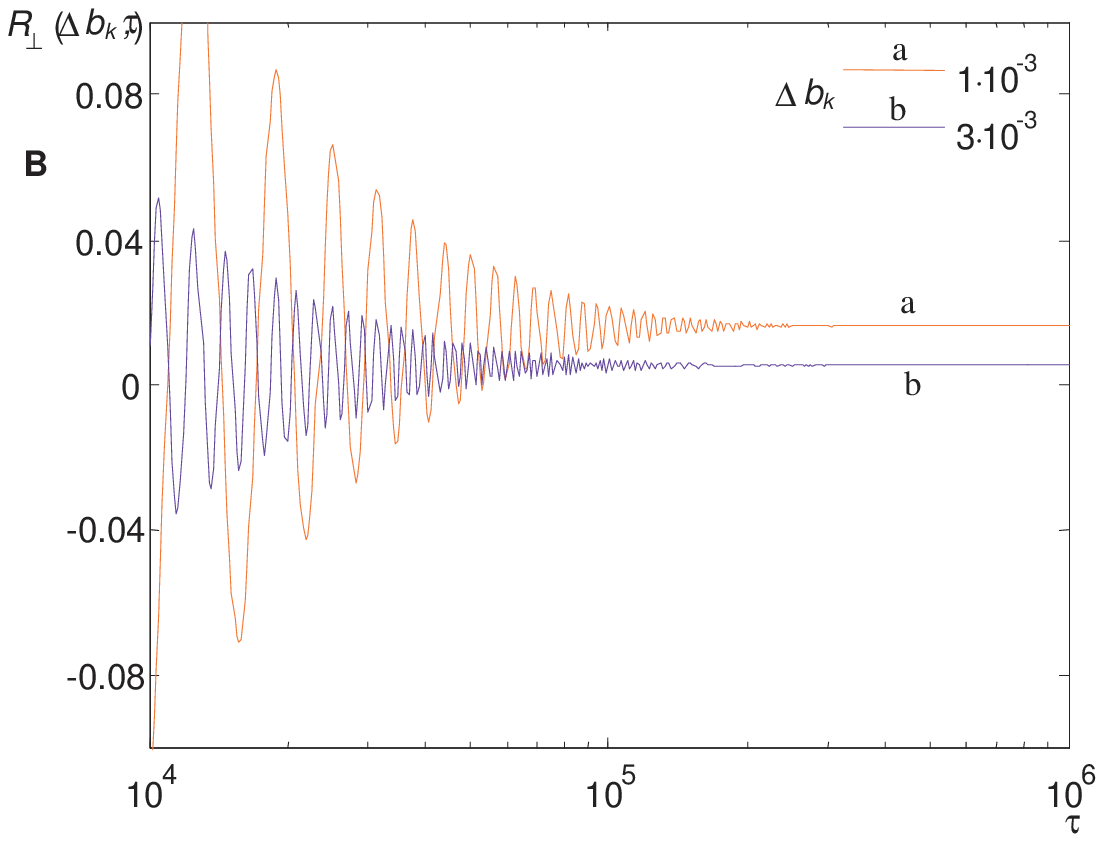}
 \end{tabular}
 \end{center}
 \caption[1]
%>>>> use \label inside caption to get Fig. number with \ref{}  Fig.~\ref{fig:2}
 {\label{fig:2} 
\textbf{A} The $\tau$-dependence of $R_{\bot}$  ($\Delta b_{k} ,\tau$) 
($\tau = t\omega_{E}$, $\omega_{E}/2\pi \sim 10^{11}\mathrm{Hz}$), 
for $b_{C}^{2} = 1/4$, $s = 10^{ - 5}$
and $\Delta b_{k} = b_{C}-b_{k} = -3 \cdot 10^{-3}$, $1 \cdot 10^{ - 3}$, $3 \cdot 10^{-3}$. 
The case \textbf{B} corresponds to the enlarged values of 
$R_{\bot} \left({\Delta b_{k},\tau}\right)$ only for $\Delta b_{k} = 1 \cdot 10^{-3}$, $3 \cdot 10^{-3}$ 
in the range of $10^{-4} < \tau < 10^{-6}$.
 }
 \end{figure}

Fig.~\ref{fig:2} gives the $\tau$-dependence of $R_{\bot} \left({\Delta b_{k} ,\tau}\right)$ 
for distinct values of parameter $\Delta b_{k} $.

%-------------
%-------------

We notice, that oscillating part of the rate of decoherence tends fast (microseconds) to constant.

Taking then $\tau \to \infty$ ($\tau \gg 1/s \sim 10^{5}$), 
we will obtain for inverse time of decoherence
\begin{equation}
\label{eq31}
1/T_{D} = \omega_{E} \Re d\gamma_{\bot} \left({k,\infty}\right)/d\tau = \omega_{E} \frac{{3a^{2}}}{{2\pi} }R_{\bot} \left({b_{k},\infty}\right) = 
\end{equation}
\nopagebreak $$ % continuation of eq2
 = \omega_{E} s\frac{{3a^{2}}}{{2\pi} }\,\,\{ \,\,\frac{{\left({\sqrt {1 + b_{C}^{2}} + b_{C} - \Delta b_{k}}\right)\left({\sqrt {b_{C}^{2} + \PISQTW} - b_{C}}\right)}}{{\Delta b_{k} \sqrt {b_{C}^{2} + \PISQTW} - b_{C} + \Delta b_{k}} } + \log\frac{{\sqrt {b_{C}^{2} + \PISQTW} - b_{C} + \Delta b_{k}} }{{\Delta b_{k}} }\}.
$$

For values 
$a^{2}\sim 10^{-6} - 10^{-8}$, 
$s \sim 10^{-5} \ll \Delta b_{k} \ll 1_{k}$, 
$\omega_{E} /2\pi \sim 10^{11}\mathrm{Hz}$
decoherence time is 
$T_{D} \sim \Delta b_{k} /\omega_{E} sa^{2}\, \mathrm{sec}$.

It follows that decoherence time, caused by one-magnon processes near 
turning points tends fast to the low value (for $\Delta b_{k} \ge 10^{-3}$ decoherence time may be from milliseconds to seconds). Note that character of nonadiabatic decoherence rate depends on the anisotropy of antiferromagnet (through parameter $b_{C}$) 
and from inhomogeneity of external field (trough parameter $\Delta b_{k}$).

The expression for frequency shift is
$$ % begining of eq32
 \Im d\gamma_{\bot} \left({k,\infty}\right)/d\tau =
$$
\nopagebreak \begin{equation}
\label{eq32}
= - \frac{{3a^{2}}}{{2\pi} }\{ \left({\sqrt {1 + b_{C}^{2}} + b_{C} - \Delta b_{k}}\right)\log\frac{{\sqrt {b_{C}^{2} + \PISQTW} - b_{C} + \Delta b_{k}} }{{\Delta b_{k}} } + \sqrt {b_{C}^{2} + \PISQTW} - b_{C} \} \approx
\end{equation}
$$ % begining of eq32
\approx - \frac{{3a^{2}}}{{2\pi} }\left({\sqrt {1 + b_{C}^{2}} + b_{C}}\right)\log\frac{{\sqrt {b_{C}^{2} + \PISQTW} - b_{C}} }{{\Delta b_{k}} }.
$$

It does not depend from magnon damping parameter.

\section{The decoherence of two qubit entangled quantum states}

The arbitrary state of pair spin-qubits in quantum register with
zero Bloch vector values
$P\left({k,\tau}\right)
= 2 \tr_{I} I_{k} \rho_{I} \left({l,k,\tau}\right)
= P\left({l,\tau}\right) = 0$
is described by the following reduced density matrix of nuclear
spin system ($\alpha,\beta = x,y,z$) (Ref.[5], \S\S 2.5-2.7):
\begin{equation}
\label{eq33}
\rho _{I} \left({l,k,\tau}\right)
 = \tr_{1,...,m \ne l,k,...N} \rho_{I} \left({1,...,l,...k,...N,\tau}\right)
 = 1/4\{ 1 + \sum\limits_{\alpha,\beta}{4G_{\alpha,\beta} \left({l,k,\tau}\right)\left({I_{l,\alpha} \otimes I_{k\beta}}\right)}\} .
\end{equation}

The non-steady pair entanglement state will be determined in the form
\begin{equation}
\label{eq34}
\rho _{I} \left({l,k,\tau}\right)
  = 1/4\,
  \{
  1 + 4 G_{z,z} \left({l,k,\tau}\right) \left({I_{l,z} I_{k,z}}\right)
    + G^{+,-}\left({l,k,\tau}\right) \left({I_{l}^{-}I_{k}^{+}}\right)
    + G^{-,+}\left({l,k,\tau}\right) \left({I_{l}^{+}I_{k}^{-}}\right)
  \} ,
\end{equation}
\noindent
where diagonal and non-diagonal elements of the matrix are defined as
\begin{equation}
\label{eq35}
G_{z,z} \left({l,k,\tau}\right)
  = 4 \tr\left({I_{l,z} I_{k,z}}\right)\,\rho _{I} \left({l,k,\tau}\right), 
\end{equation}
\nopagebreak $$ % continuation of eq35
G^{+,-} \left({l,k,\tau}\right)
  = \left({G^{-,+} \left({l,k,\tau}\right)}\right)^{*}
  = 4 \tr\left({I_{l}^{+} I_{k}^{-} }\right)\,\,\rho _{I} \left({l,k,\tau}\right).
$$

In the interaction representation relative Hamiltonian
\begin{equation}
\label{eq36}
h_{0} \left({l,k}\right) = h_{S} - \left({\left({\omega_{I}\left({l}\right) - a/2}\right)\;I_{z} \left({l}\right) + \left({\omega_{I} \left({k}\right) - a/2}\right)\;I_{z} \left({k}\right)}\right)
\end{equation}
\noindent
the reduced density matrix has the form
\begin{equation}
\label{eq37}
\rho _{in} \left({l,k,}\right) = 
\exp\left({ih_{0} \left({l,k}\right)\tau}\right)\rho \left({l,k,\tau}\right)\exp\left({-ih_{0} \left({l,k}\right)\tau}\right).
\end{equation}

Let there be the pure triplet entangled state of two removed spins $l$ and $k$ with zero total 
$z$-projection $I = 1$, $M = 0$, 
which belongs to the same sublattice, realized by the certain external action in the initial moment 
$\tau = 0$. It will be described by reducer state vector 
$|1,\,0\rangle = \sqrt{1/2} \left({| \uparrow \downarrow \rangle + | \downarrow \uparrow \rangle}\right)$ 
or following density matrix:
\begin{equation}
\label{eq38}
\rho _{I} \left({l,k,0}\right) = |1,\,0\rangle \langle 1,\,0| = \frac{{1}}{{2}}\;\left| {\;{\begin{array}{*{20}c}
 {0} \hfill & {0} \hfill & {0} \hfill & {0} \hfill \\
 {0} \hfill & {1} \hfill & {1} \hfill & {0} \hfill \\
 {0} \hfill & {1} \hfill & {1} \hfill & {0} \hfill \\
 {0} \hfill & {0} \hfill & {0} \hfill & {0} \hfill \\
\end{array}} }\right| = 1/4\{ 1 - 4\left({I_{l,z} I_{k,z}}\right) + 2\left({I_{l}^{-} I_{k}^{+} }\right) + 2\left({I_{l}^{+} I_{k}^{-}}\right)\} ,
\end{equation}
\noindent
where 
$G_{z,z} \left({l,k,0}\right) = - 1$,
$G^{+,-} \left({l,k,0}\right) = G^{-,+} \left({l,k,0}\right) = 2$. 
The concurrence of this entangled state has maximum value $C = 1$.

The evolution of such two-qubit state is due to the qubit interaction with magnons
which in considered model is described by the following perturbation Hamiltonian 
in interaction representation
\begin{equation}
\label{eq39}
\Delta h_{IS} \left({l,k,\tau}\right) = 
\Delta h_{IS} \left({l,\tau}\right) + \Delta h_{IS} \left({k,\tau}\right) = 
\end{equation}
\nopagebreak $$ % continuation of eq39
 = \exp\left({ih_{0} \left({l,k}\right)\tau}\right)\left({\Delta h_{IS} \left({l}\right) + \Delta h_{IS} \left({k}\right)}\right)\exp\left({-ih_{0} \left({l,k}\right)}\right)\tau ,
$$
\noindent
where values $\Delta h_{IS} \left({l,\tau}\right)$, $\Delta h_{IS} \left({k,\tau}\right)$ 
are determined above by expression (\ref{eq8}).

Let us suppose now (as previously in Sec.3), 
that the interaction of nuclear spins in ground coherent state (\ref{eq38}) 
with electron system of the antiferromagnet is turning on at initial moment $\tau = 0$
when non-disturbed density matrix has the form of direct product: 
$\rho \left({0}\right) = \rho _{I} \left({l,k,0}\right)\rho _{S} \left({0}\right) = \rho _{I} \left({l,k,0}\right)|0\rangle \langle 0|$.

We will write next the equation for density matrix in the interaction representation 
$\rho_{in} \left({l,k,\tau}\right)$:
\begin{equation}
\label{eq40}
i\partial \rho_{in} \left({l,k,\tau}\right)/\partial \tau = \left[ {\Delta h_{IS} \left({l,k,\tau}\right),\,\,\rho _{in} \left({l,k,\tau}\right)}\right].
\end{equation}

From Eq.(\ref{eq40}) for density matrix in the second theory of perturbation 
theory it is follow the equation
\begin{equation}
\label{eq41}
i\partial \rho _{in} \left({l,k,\tau}\right)/\partial \tau = \left[ {\Delta h_{IS} \left({l,k,\tau}\right),\,\rho _{in} \left({l,k,0}\right)|0\rangle \langle 0|}\right] - 
\end{equation}
\nopagebreak $$ % continuation of eq2 
- i\int\limits_{0}^{\tau} {\left[ {\Delta h_{IS} \left({l,k,\tau}\right),\,\,\left[ {\Delta h_{IS} \left({l,k,{\tau}'}\right),\,\rho _{I} \left({l,k,0}\right)|0\rangle \langle 0|}\right]}\right]} \,d{\tau}'. 
$$                                                

By using Eq.(\ref{eq41}) and making the cyclic permutation, we will obtain 
for the elements of reduced density matrix (\ref{eq34}) the equations:
\begin{equation}
\label{eq42}
\partial \left({G_{z,z} \left({l,k,\tau}\right)}\right)/\partial \tau = 4 \tr_{I} \left({I_{l,z} I_{k,z} }\right)\partial \rho _{in} \left({l,k,\tau}\right)/\partial \tau \approx \\ 
\end{equation}
\nopagebreak $$ % continuation of eq34
 \approx - 4 \tr_{I} \,\int\limits_{0}^{\tau} {\langle 0|\left[ {\left[ {\left({I_{l,z} I_{k,z}}\right),\Delta h_{IS} \left({l,k,\tau}\right)}\right],\,\,\,\Delta h_{IS} \left({l,k,{\tau} '}\right)}\right]\rho _{I} \left({l,k,0}\right)\,|0\rangle} d{\tau} ', 
$$
\noindent
and
\begin{equation}
\label{eq43}
\partial G^{+,-} \left({l,k,\tau}\right)/\partial \tau
  = 4 \tr\,\left({I_{l}^{+} I_{k}^{-} }\right)\partial \,\rho _{in} \left({l,k,\tau}\right)/\partial \tau \approx 
\end{equation}
\nopagebreak $$ % continuation of eq2
\approx - 4 \tr\,\int\limits_{0}^{\tau} {\langle 0|\left[ {\left[ {\left({I_{l}^{+} I_{k}^{-} }\right),\Delta h_{IS} \left({l,k,\tau}\right)}\right],\,\Delta h_{IS} \left({l,k,{\tau}'}\right)}\right]\rho_{I} \left({l,k,0}\right)|0\rangle} \,d{\tau}'. 
$$

Let us take into account Eq.(\ref{eq38}), define the tensor longitudinal component in the form
\begin{equation}
\label{eq44}
G_{z,z} \left({l,k,\tau}\right)
  = - \exp\left[ {-\gamma_{\parallel} \left({l,k,\tau}\right)}\right]
\end{equation}
\noindent
and the tensor transverse component in the form
\begin{equation}
\label{eq45}
G^{+,-} \left({l,k,\tau}\right)
  = 2 \exp\left[ {i\left({\omega_{I} \left({l}\right) - \omega_{I} \left({k}\right)}\right)\tau - \gamma_{\bot} \left({l,k,\tau}\right)}\right].
\end{equation}

To obtain the expression for the rates of pair relaxation of longitudinal and 
transverse components in the context of second order of perturbation theory, we will write:
\begin{equation}
\label{eq46}
\partial G_{z,z} \left({l,k,\tau}\right)/\partial \tau = d\gamma_{\parallel} \left({l,k,\tau}\right)/d\tau = d\gamma_{\parallel}^{*} \left({l,k,\tau}\right)/d\tau =
\end{equation}
\nopagebreak $$ % continuation of eq46
= - 4      \tr_{I} \int\limits_{0}^{\tau} {\langle 0|\left[ {\left[ {\left({I_{l,z} I_{k,z}}\right),\Delta h_{IS} \left({l,k,\tau}\right)}\right],\Delta h_{IS} \left({l,k,{\tau}'}\right)}\right]\left({-\left({I_{l,z} I_{k,z}}\right) + 1/2\left({I_{l}^{+} I_{k}^{-} }\right) + 1/2\left({I_{l}^{-} I_{k}^{+} }\right)}\right)
|0\rangle d{\tau} '} =
$$
\nopagebreak $$ % continuation of eq49
= d\gamma_{\parallel} \left({l,\tau}\right)/d\tau + d\gamma_{\parallel} \left({k,\tau}\right)/d\tau + d\tilde {\Gamma} _{\parallel} \left({k,l - k,\tau}\right)/d\tau 
$$
\noindent
and
\begin{equation}
\label{eq47}
\Re\partial G^{+, -} \left({l,k,\tau}\right)/\partial \tau = - 2\Re d\gamma_{\bot} \left({l,k,\tau}\right)/d\tau = \\ 
\end{equation}
\nopagebreak $$ % continuation of eq47
= - 4\Re \tr_{I} \int\limits_{0}^{\tau} {\langle 0|\left[ {\left[ {\left({I_{l}^{+} I_{k}^{-}}\right),\Delta h_{IS} \left({l,k,\tau}\right)}\right],\Delta h_{IS} \left({l,k,{\tau}'}\right)}\right]} \left({-\left({I_{l,z} I_{k,z}}\right) + 1/2\left({I_{l}^{+} I_{k}^{-} }\right) + 1/2\left({I_{l}^{-} I_{k}^{+} }\right)}\right)
|0\rangle d{\tau}' = \\
$$ 
\nopagebreak $$ % continuation of eq47
 = - 2\Re d\gamma_{\bot} \left({l,\tau}\right)/d\tau - 2\Re d\gamma_{\bot} \left({k,\tau}\right)/d\tau - 2\Re d\tilde {\Gamma} _{\bot} \left({k,l - k,\tau}\right)/d\tau ,
$$ 
\noindent
where the correlation part of longitudinal two spin relaxation rate 
$$ % begining of eq48
 d\tilde {\Gamma} _{\parallel} \left({k,l - k,\tau}\right)/d\tau =
$$ 
\nopagebreak \begin{equation}
\label{eq48}
= - 4 \tr_{I} \,\int\limits_{0}^{\tau} {\langle 0|\left[ {\left[ {I_{l,z} ,\Delta h_{IS} \left({l,\tau}\right)}\right]I_{k,z} ,\,\,\,\Delta h_{IS} \left({k,{\tau} '}\right)}\right]} \left({\left({I_{l}^{+} I_{k}^{-}}\right) + \left({I_{l}^{-} I_{k}^{+} }\right)}\right)\,|0\rangle \,d{\tau} ' =
\end{equation}
\nopagebreak $$ % continuation of eq48
 = - \frac{{a^{2}}}{{4}}4\Re\int\limits_{0}^{\tau} {\langle 0|\left[ {S_{k}^{+} \left({{\tau} '}\right),S_{l}^{-} \left({\tau}\right)}\right]|0\rangle} \,d{\tau} '. 
$$

Consider next the correlation part of transverse two spin decoherence rate
$$ % begining of eq49
\Re d\tilde {\Gamma} _{\bot} \left({k,l - k,\tau}\right)/d\tau =
$$
\nopagebreak $$ % continuation of eq49
 = - 2\Re\tr_{I} \,\int\limits_{0}^{\tau} {\langle 0|\left[ {\left[ {I_{l}^{+} ,\Delta h_{IS} \left({l,\tau}\right)}\right]I_{k}^{-} ,\,\,\,\Delta h_{IS} \left({k,{\tau} '}\right)}\right]\left({I_{l,z} I_{k,z}}\right)} \,\, +
$$
\nopagebreak \begin{equation}
\label{eq49}
 + \left[ {\left[ {I_{k}^{-} ,\Delta h_{IS} \left({k,\tau}\right)}\right]I_{l}^{+} ,\,\,\,\Delta h_{IS} \left({l,{\tau} '}\right)}\right]\left({I_{l,z} I_{k,z}}\right)|0\rangle d{\tau} ' =
\end{equation}
\nopagebreak $$ % continuation of eq49
 \, = - \frac{{a^{2}}}{{4}}2\Re\int\limits_{0}^{\tau} {\langle 0|\left[ {S_{k}^{-} \left({{\tau} '}\right),S_{l}^{+} \left({\tau}\right)}\right] + \left[ {S_{k}^{+} \left({{\tau}'}\right),S_{l}^{-} \left({\tau}\right)}\right]|0\rangle} d{\tau}' =
$$ 
\nopagebreak $$ % continuation of eq49
= - \frac{{a^{2}}}{{4}}4\Re\int\limits_{0}^{\tau} {\langle 0|\left[ {S_{k}^{+} \left({{\tau} '}\right),S_{l}^{-} \left({\tau}\right)}\right]|0\rangle} d{\tau}',
$$
\noindent
that is the correlation part of decoherence rate is equal 
to the correlation part of longitudinal relaxation rate.

Let us rewrite next the expression (\ref{eq49}) 
similarly as it would made in Sec.3, accounting only 
low magnon excitation mode with energy $E_{-}$:
$$ % begining of eq50
\Re d\tilde {\Gamma} _{\bot} \left({k,l-k,\tau}\right)/d\tau \approx 
$$
\nopagebreak $$ % continuation of eq50
\approx \frac{{a^{2}}}{{\left({4\pi}\right)^{2}}}2\Re\int {\int\limits_{o}^{\tau} {u^{*} \left({q_{\bot} ,E}\right)u\left({{q}'_{x} ,q_{y} ,E}\right)\exp\left({-\left({iE_{-} - s}\right)\left({{\tau} ' - \tau}\right)}\right)\exp\left[ {i\left({q_{x} k - {q}'_{x} l}\right)}\right]dEdq_{\bot} d{q}'_{x} d{\tau} '}} =$$
\nopagebreak \begin{equation}
\label{eq50}
 = \frac{{a^{2}}}{{\left({4\pi}\right)^{2}}}\int {\frac{{\sqrt {1 + b_{C}^{2}} + E\left({q_{\bot} }\right)}}{{E\left({q_{\bot} }\right)}} \cdot} \{ \,\frac{{-\left({E\left({q_{\bot} }\right) - b_{\left({l + k}\right)/2}}\right)\sin\left({q_{x} \left({l-k}\right)}\right) + s \cos\left({q_{x} \left({l-k}\right)}\right)}}{{\left({E\left({q_{\bot} }\right) - b_{\left({l + k}\right)/2}}\right)^{2} + s^{2}}} + 
\end{equation}
\nopagebreak $$ % continuation of eq50
 + \frac{{\left[ {\left({E\left({q_{\bot} }\right) - b_{\left({l + k}\right)/2}}\right)\sin\left({\left({E\left({q_{\bot} }\right) - b_{\left({l + k}\right)/2}}\right)\tau - q_{x} \left({l-k}\right)}\right) - s \cos\left({\left({E\left({q_{\bot} }\right) - b_{\left({l + k}\right)/2}}\right)\tau - q_{x} \left({l-k}\right)}\right)}\right]\exp\left({-s\tau}\right)}}{{\left({E\left({q_{\bot} }\right) - b_{\left({l + k}\right)/2}}\right)^{2} + s^{2}}}\} dq_{\bot} . 
$$

\begin{figure}
 \begin{center}
 \begin{tabular}{c}
 \includegraphics{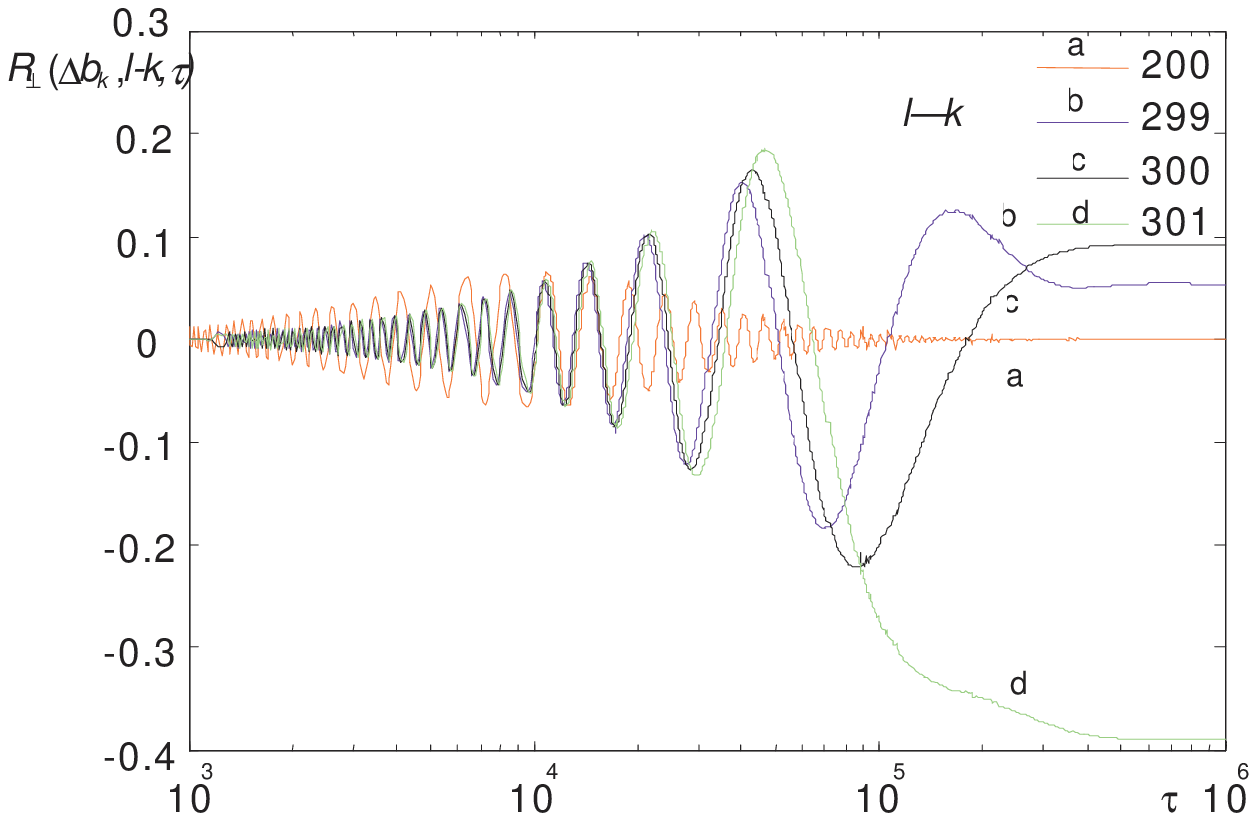}
 \end{tabular}
 \end{center}
 \caption[1]
%>>>> use \label inside caption to get Fig. number with \ref{}  Fig.~\ref{fig:3}
 {\label{fig:3} 
The $\tau$-dependence of decoherence rate $R_{\bot} \left({\Delta b_{k},l-k,\tau}\right)$ 
for values $\Delta b_{k} = 3.10^{-3},$ $s = 10^{-5}$, $l-k = 200, 299, 300, 301$.
 }
 \end{figure}

Taking next in to account that 
$\int\limits_{0}^{2\pi} {\sin\left[ {q_{\bot} \cos \varphi \left({l-k}\right)}\right]}d\varphi = 0$
and introducing again the variable
$\xi = E\left({q_{\bot} }\right) - b_{C}$,
we will write
$E\left({q_{\bot} }\right) - b_{\left({l + k}\right)/2} = E\left({q_{\bot} }\right) - b_{k} - g\left({l-k}\right)/2 = \xi + \Delta b_{k} - g\left({l-k}\right)/2$ and obtain (Fig.~\ref{fig:3})
\begin{equation}
\label{eq51}
\frac{{2\pi }}{{3a^{2}}} \Re d\tilde{\Gamma}_{\bot} \left({k,l-k,\tau}\right)/d\tau
 \approx R_{\bot} \left({\Delta b_{k} ,l - k,\tau}\right) = \\
\end{equation}
\nopagebreak $$ % continuation of eq50
 = \int\limits_{0}^{\sqrt {b_{C}^{2} + \PISQTW} - b_{C}} {\left({\sqrt {1 + b_{C}^{2}} + b_{C} + \xi}\right) Y\left({\xi +\Delta b_{k} - g\left({l-k}\right)/2,\tau}\right) J_{0} \left({\sqrt{12\left[ {\left({b_{C} + \xi}\right)^{2} - b_{C}^{2}}\right]} \left({l-k}\right)}\right)d\xi} .
$$

%-------------
%-------------

The decoherence rates of entangled qubit pair are due to decoherence of one spin states 
$l$, $k$ and also to the correlation between nuclear spins $l$, $k$. 
The initial diagonal and non-diagonal elements of density matrix
$1 + G_{z,z} \left({l,k,\tau}\right)$
and $G^{+,-}\left({l,k,\tau}\right)$ are decreased with full rates
$\Re d\Gamma_{\parallel} \left({l,k,\tau}\right)/d\tau$ and
$\Re d\Gamma_{\bot} \left({l,k,\tau}\right)/d\tau$.
Note that the asymptotic value for correlation part of
decoherence rate (Fig.~\ref{fig:3}) may periodically change sign, 
if $\Delta b_{k} - g\left({l-k}\right)/2 < 0$, that is for states after ``turning point''. 
Likewise, the indirect interspin interaction is described by precisely the same oscillating 
sign-changing function after ``turning point'' (see Eq. (74) in Ref.[4]).

The concurrence for entangled two-qubit state can be obtained by using the Wootter formula (Ref.[6])
\begin{equation}
\label{eq52}
C\left({l,k,\tau}\right) = 
1/2 \max\{ |G^{+,-} \left({l,k,\tau}\right)| - \left({1 + G_{z,z} \left({l,k,\tau}\right)}\right);\,0\} .
\end{equation}

Taking in to account that
%%\begin{equation}
%%\label{eq52a}
$\Re\gamma_{\bot} \left({k,\tau}\right)
  = \gamma_{\parallel} \left({k,\tau}\right)$,
$|G^{+,-} \left({l,k,\tau}\right)|
   \approx 2\exp\left({-\Re\gamma_{\bot} \left({l,k,\tau}\right)}\right)$,
$G_{z,z} \left({l,k,\tau}\right)
  \approx - \exp\left({-\gamma_{\parallel} \left({l,k,\tau}\right)}\right)$ 
%%\end{equation}
and also Eqs (\ref{eq44}), (\ref{eq45}), (\ref{eq47}), we will obtain
\begin{equation}
\label{eq53}
C\left({l,k,\tau}\right) = 1/2\{ 3 \exp\left({-\Re\gamma_{\bot} \left({l,\tau}\right) -\Re\gamma_{\bot} \left({k,\tau}\right) - \Re\tilde {\Gamma} _{\bot} \left({k,l-k,\tau}\right)}\right) - 1\} .
\end{equation}

For the concurrence-damping rate we will then write
\begin{equation}
\label{eq54}
 dC\left({l,k,\tau}\right)/d\tau
   = - 3/2d\left({\Re\gamma_{\bot} \left({l,\tau}\right) + \Re\gamma_{\bot} \left({k,\tau}\right) + \Re\tilde {\Gamma} _{\bot} \left({k,l - k,\tau}\right)}\right)d\tau  =
\end{equation}
\nopagebreak $$ % continuation of eq54
 = - \frac{{9a^{2}}}{{2\pi} }\left({R_{\bot} \left({\Delta b_{l} ,\tau}\right) + R_{\bot} \left({\Delta b_{k} ,\tau}\right) + R_{\bot} \left({\Delta b_{k} ,l - k,\tau}\right)}\right).
$$

Note, that for $\Delta b_{k} > 0$ parameter $\Delta b_{l} = \Delta b_{k} - g\left({l-k}\right)$ can change the sign. The expression $R_{\bot} \left({\Delta b_{l} ,\tau}\right)$ takes the form
\begin{equation}
\label{eq55}
R_{\bot} \left({\Delta b_{l} ,\tau}\right) 
= R_{\bot} \left({\Delta b_{k} - g\left({l-k}\right),\tau}\right) 
= \int\limits_{0}^{\sqrt {b_{C}^{2} + \PISQTW} - b_{C}} {\left({\sqrt {1 + b_{C}^{2}} + b_{C} + \xi}\right) Y\left({\xi + \Delta b_{k} - g\left({l-k}\right),\tau}\right) d\xi} 
\end{equation}

\begin{figure}
 \begin{center}
 \begin{tabular}{c}
 \includegraphics{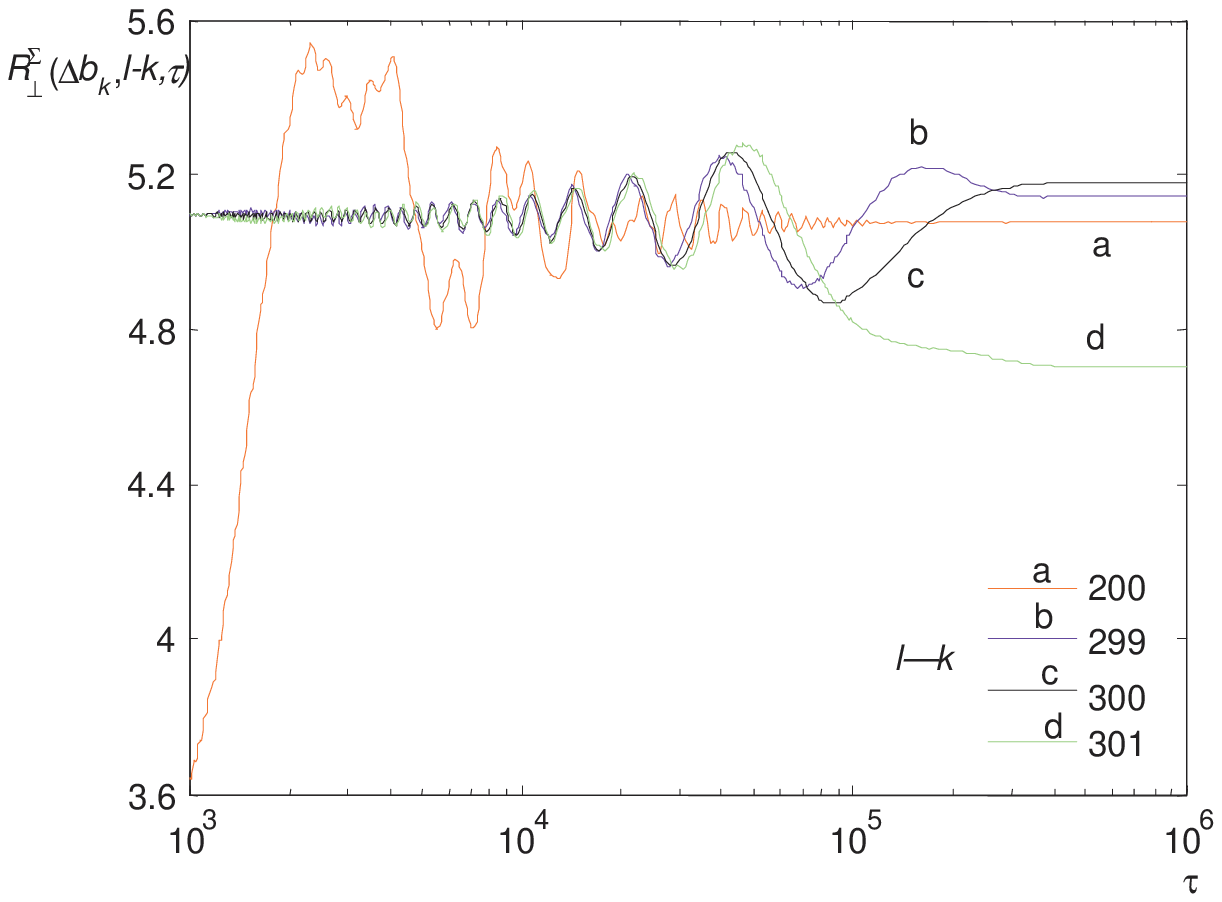}
 \end{tabular}
 \end{center}
 \caption[1]
%>>>> use \label inside caption to get Fig. number with \ref{}  Fig.~\ref{fig:4}
 {\label{fig:4} 
 The $\tau - $dependence of $\,R_{\bot} ^{\Sigma} \left({\Delta b_{k} ,l - k,\tau}\right)$ for values $\Delta b_{k} = 3.10^{-3},\,\,\,\,s = 10^{-5},\,\,\,\,\,\,\,l - k = 200,\,\,299,\,\,300,\,\,301.$
 }
 \end{figure}

Finally, for the value of concurrence damping rate we will obtain
\begin{equation}
\label{eq56}
dC\left({l,k,\tau}\right)/d\tau = - \frac{{9a^{2}}}{{2\pi} }R_{\bot} ^{\Sigma} 
\left({\Delta b_{k} ,l - k,\tau}\right),
\end{equation}
\noindent
where (Fig.~\ref{fig:4})
$$ % beginning of eq57
 R_{\bot} ^{\Sigma} \left({\Delta b_{k} ,l - k,\tau}\right)
 = R_{\bot} \left({\Delta b_{k} ,\tau}\right) + R_{\bot} \left({\Delta b_{l} ,\tau}\right) + R_{\bot} \left({\Delta b_{k} ,l - k,\tau}\right) = 
$$
\nopagebreak \begin{equation}
\label{eq57}
 = \int\limits_{0}^{\sqrt {b_{C}^{2} + \PISQTW} - b_{C}} {\left({\sqrt {1 + b_{C}^{2}} + b_{C} + \xi}\right) \cdot \left[ {Y\left({\xi + \Delta b_{k} ,\tau}\right)} + Y\left({\xi + \Delta b_{k} - g\left({l-k}\right),\tau}\right) \right.} + \\
\end{equation}
\nopagebreak $$ % continuation of eq57
 + \left.{
 Y\left({\xi + \Delta b_{k} - g\left({l-k}\right)/2,\tau}\right)\, J_{0} \left({\sqrt {12 \left[ {\left({b_{C} + \xi}\right)^{2} - b_{C}^{2} }\right] } \left({l-k}\right)}\right)}\right] d\xi .
$$

%-------------
%-------------

The concurrence damping rate tends at $\tau \to \infty$ to positive value:
$$ % beginning of eq58
 R_{\bot} ^{\Sigma} \left({\Delta b_{k} ,l - k,\infty}\right)\,\, = s\int\limits_{0}^{\sqrt {b_{C}^{2} + \PISQTW} - b_{C}} {\left({\sqrt {1 + b_{C}^{2}} + b_{C} + \xi}\right)\left[ {\frac{{1}}{{\left({\xi + \Delta b_{k}}\right)^{2}}}\, + \frac{{1}}{{\left({\xi + \Delta b_{k} - g\left({l-k}\right)}\right)^{2}}}\,\, +}\right.}
$$
\nopagebreak \begin{equation}
\label{eq58}
 \left .{
 + \frac{{J_{0} \left({\sqrt {12\left[ {\left({b_{C} + \xi}\right)^{2} - b_{C}^{2}}\right]} \left({l-k}\right)}\right)}}{{\left({\xi + \Delta b_{k} - g\left({l-k}\right)/2}\right)^{2}}}\,\,}\right]\,d\xi > 0.
\end{equation}

\section{Adiabatic decoherence caused by interaction with nuclear spins of random 
distributed isotopes substituting the basic isotopes in antiferromagnetic structure.}

Between other mechanisms of decoherence it should be pointed out the adiabatic mechanism
that is determined by magnetic interaction of nuclear spins-qubits with electron and nuclear
spins of impurity atoms, which play here role of an environment (Ref.[5], \S 5.4).

Interaction of nuclear spins with magnetic moments of impurity paramagnetic atoms is of
no concern as compared to the interaction of nuclear spins with electron spins of own atoms.
This mechanism is largely suppressed for a high degree of electron spin polarization
(at $B/T > 30\,\mathrm{T}/\mathrm{K}$).

Let us consider here a decoherence model, where nuclear spin-qubit interact
with nuclear magnetic moments of randomly distributed impurity isotopes in
basic nuclear spin-free antiferromagnet.

Hamiltonian of dipole-dipole magnetic interaction of considered nuclear spins
has the following form:

\begin{equation}
\label{eq59}
H_{I,I_{imp}} = \hbar \sum\limits_{i,\alpha \beta} ^{N} {D_{\alpha \beta} \left({r_{i}}\right)} \;I_{\alpha} I_{\beta ,imp} \left({r_{i} ,t}\right),
\end{equation}
\noindent
where
\begin{equation}
\label{eq60}
D_{\alpha \beta} \left({r_{i}}\right)\frac{{\mu _{0}} }{{4\pi }}\frac{{\gamma_{I} \gamma_{I,imp} \hbar} }{{r_{i}^{3}} }\left({\delta _{\alpha \beta} - \frac{{3r_{i\alpha} r_{i\beta} } }{{r_{i}^{2}} }}\right) \quad ,
\end{equation}
$\gamma_{I} /2\pi$, $\gamma_{I,imp} /2\pi$ 
are giromagnetic ratio of quantum register nuclear spin-qubit and of impurity isotope nuclear spin, $\mathbf{r}_{i}$ is radius-vector of distance from the position of nuclear spin-qubit to position of $i$-th impurity nuclear spin, $\gamma_{I,imp} \hbar I_{\beta ,imp} \left({r_{i} ,t}\right)$ is fluctuating magnetic moment of impurity isotope, which produce the random local field:
\begin{equation}
\label{eq61}
\Delta B_{\alpha} \left({t}\right) = -\sum\limits_{i,\beta}^{N}{D_{\alpha \beta} \left({r_{i}}\right)\left({I_{\beta ,imp} \left({r_{i} ,t}\right) - \langle I_{\beta ,imp} \left({r_{i}}\right)\rangle }\right)/\gamma_{I} } .
\end{equation}

The mean value $\langle \gamma_{I} B_{\alpha} \left({t}\right)\rangle = \sum\limits_{i,\beta} ^{N} {D_{\alpha \beta} \left({r_{i}}\right)} \langle I_{\beta ,imp} \left({r_{i}}\right)\rangle $ determines the shift of qubit resonance frequency.

The correlation function for random modulation of nuclear spin resonance frequency is determined by the following expression

\begin{equation}
\label{eq62}
\left\langle {\Delta \omega \left({\tau}\right)\Delta \omega \left({0}\right)}\right\rangle = C_{I,imp} \int {\sum\limits_{\beta} {D_{z\beta }^{2} \left({r}\right)\left({\langle I_{\beta ,imp} \left({r,\tau}\right)I_{\beta,imp} \left({r,0}\right)\rangle - \langle I_{\beta,imp} \left({r}\right)\rangle ^{2}}\right)} dr} ,
\end{equation}
\noindent
where $C_{I,imp}$ is concentration of impurity isotopes .

The expression for correlation function in considered case of adiabatic decoherence takes the form
\begin{equation}
\label{eq63}
\left\langle {\Delta \omega \left({\tau}\right)\Delta \omega \left({0}\right)}\right\rangle = \langle \Delta \omega ^{2}\rangle \exp\left({-t/T_{\parallel,imp}}\right),.
\end{equation}
\noindent
where quadratic mean value of modulation frequency

\begin{equation}
\label{eq64}
\langle \Delta \omega ^{2}\rangle = C_{I,imp} \left({\frac{{\mu _{0} }}{{4\pi} }\gamma_{I} \gamma_{I,imp} \hbar}\right)^{2}\frac{{16\pi}}{{15a^{3}}}\left({\langle I_{z,imp}^{2} \rangle - \langle I_{z,imp} \rangle ^{2}}\right),
\end{equation}

\begin{equation}
\label{eq65}
\langle I_{z,imp}^{2} \rangle - \langle I_{z,imp} \rangle ^{2} = \left({1 - \th^{2}\left({\gamma_{I,imp} B\hbar /2kT_{I}}\right)}\right)/4,
\end{equation}
\noindent 
$a$ is minimal distance to impurity nuclear spin, with is of order of lattice constant $\sim$ 1 nm.

If it is believed that $T_{\parallel,imp} \sim 10^{4}\;\mathrm{sec} \gg T_{D}$ 
(the so-called condition of rigid lattice) and that $T_{D} > 1\;\mathrm{sec}$, 
for the determination of allowable isotope concentration we will obtain the condition
\begin{equation}
\label{eq66}
1/T_{D}^{2} \approx \langle \Delta \omega ^{2}\rangle = C_{I,imp} \left({\frac{{\mu_{0}} }{{4\pi} }\gamma_{I} \gamma_{I,imp} \hbar}\right)^{2}\frac{{4\pi} }{{15a^{3}}}\left({1 - \th^{2}\left({|\gamma_{I,imp} |B\hbar /2kT_{I}}\right)}\right) < 1\;\mathrm{sec}^{-2}.
\end{equation}

For values $B/T > 30\;\mathrm{T}/\mathrm{K}$
we will obtain for allowable concentration
of impurity isotopes the highly rigid condition
$C_{I,imp} < 10^{15}\,\mathrm{cm}^{-3}$,
$\left({\sim 10^{-5}\,\%}\right)$.
However, if temperature $T_{I}$ 
for nuclear spins of impurity isotopes corresponds to value,
for which almost full nuclear spin polarization takes place
$|\gamma_{I,imp} |B\hbar /kT_{I} > 1$,
that is $T_{I} < 1\, \mathrm{mK}$,
so the allowable concentration in isotope-pure antiferromagnet will
$C_{I,imp} \% < 4.5 \cdot 10^{-2}\% $.
It will rapidly increase on further lowering of nuclear spin temperature.

Eventually the suppressing of this mechanism calls for appropriate cleaning of substrate from impurity atoms and using very low spin temperatures for nuclear spins.

\section{The encoded DFS (Decoherence-Free Subspaces) logical qubits are constructed on clusters of the four-physical qubits}

Let us consider here two encoded DFS logical qubits
$|0_{L} \rangle$ and $|1_{L} \rangle$
with zero total angular momentum $J = 0$, $m_{J} = 0$,
which states are constructed on cluster of four states of physical qubits:
\begin{equation}
\label{eq67}
 |0_{L} \rangle = 1/2\left[ {\left({|01\rangle - |10\rangle}\right) \otimes \left({|01\rangle - |10\rangle}\right)}\right] 
\end{equation}
\nopagebreak $$ % continuation of eq67
 |1_{L} \rangle = 1/\sqrt {3} \,\left[ {|11\rangle \otimes |00\rangle + |00\rangle \otimes |11\rangle - 1/2\left({|01\rangle + |10\rangle}\right) \otimes \left({|01\rangle + |10\rangle}\right)}\right].
$$

They form two dimensional subspace for quantum operation.

As is shown in Refs. [7-9] this subspace represent, as it is called,
the strong collective decoherence free subspace (DFS),
if the states of four physical qubits are the eigenstates
of Hamiltonian with antiferromagnetic interaction $\left({\Delta > 0}\right)$
of XXX type in the absent of external field
\begin{equation}
\label{eq68}
h_{xxx}^{4} = \Delta /2\sum\limits_{k \ne l}^{4} {I\left({k}\right)I\left({l}\right)}
= \Delta /2\{ \sum\limits_{k}^{4} {I\left({k}\right)} \sum\limits_{l}^{4} {I\left({l}\right)} - 3 \mathbf{1}\} ,
\end{equation}
\noindent
where $\mathbf{1}$ is unit four-dimensional matrix 
(here symbol $\otimes$ is omitted) and is not relevant.
It has eigenvalues $\Delta /2\,J\left({J + 1}\right)$
with $J = 0,\,1,\,2$
The DFS states (\ref{eq67}) are the lowest state which identical to ground state
of Hamiltonian (\ref{eq68}) with $J = 0$, $m = 0$
and has two-fold degeneracy.

We will use now the before obtained expression for effective
Hamiltonian of two nuclear spins, belonging to common sublattice
in quantum register ([4], Eq.(A2.8)) with interaction of XX0 or two-dimensional isotropic type:
\begin{equation}
\label{eq69}
 h_{xxo} = - \sum\limits_{j = k,l} {\{ \left[ {\omega_{I} \left({j}\right) - a/2 - W\left({j}\right)}\right]I_{z} \left({j}\right) + \,\,U\left({j,j}\right)/2} - 
\end{equation}
\nopagebreak $$ % continuation of eq69
 - U\left({k,l}\right)\left[ {I^{-} \left({k}\right)I^{+} \left({l}\right) + I^{+} \left({k}\right)I^{-} \left({l}\right)}\right] \}
$$

We write next the total nuclear spin Hamiltonian of quantum register neglecting the terms
$W\left({j}\right) \approx U\left({j,j}\right) = Const.$
of order of $a^{2}$
and take in to account that the difference of $\omega_{I} \left({k}\right) - \omega_{I} \left({l}\right) = \gamma_{I} /\gamma_{S} g\left({k-l}\right) \ll a\sim 10^{-3}$ for 
$\left({k-l}\right) < 1/{g}\sim 10^{5}$):
\begin{equation}
\label{eq70}
h_{II} = - \left({\omega_{I} - a/2}\right)\,\sum\limits_{k} {I_{z} \left({k}\right)} \, - \sum\limits_{k \ne l} {U\left({k,l}\right)\left[ {I^{-}\left({k}\right)I^{+} \left({l}\right) + I^{+} \left({k}\right)I^{-} \left({l}\right)}\right]} .
\end{equation}

Our Hamiltonian (\ref{eq70}), written for four qubits, differs from Eq.(\ref{eq68})
in that it has identical values of interaction parameter $\Delta$
for different qubit pairs and it is free from the term
$\sum\limits_{k \ne l}^{4} {I_{z} \left({k}\right)I_{z} \left({l}\right)}$.
In addition, the Hamiltonian (\ref{eq70}) has operator $\sum\limits_{k} {I_{z} \left({k}\right)}$.

In Ref.[4] it was obtained that the dependence of indirect interspin interaction
$U\left({k,l}\right)$
from distance for
$\left({l-k}\right) > 2\Delta b_{k} /g = 2\left({b_{C} - b - gk}\right)/g$
takes oscillating character with quasi-period
which gradually decreased with increasing of the distant $\left({l-k}\right)$.
In this case indirect interaction periodically change the sign.
Consequently, it is possible to choose the nuclear spin position in such a way that the value of interaction
$U\left({k,l}\right) = - U < 0$
would be the same for all four qubit position in the nuclear spin chain
considered as a quantum register.

$$ % beginning of eq1
 h_{xxo}^{4} =
   - \left({\omega_{I}-a/2}\right)\sum\limits_{k=0}^{4} {I_{z}\left({k}\right)} + U\{ 1/2\left[ {\sum\limits_{k=0}^{4} {I^{-}\left({k}\right)} \sum\limits_{l=0}^{4}{I^{+}\left({l}\right)} +       
 \sum\limits_{k=0}^{4} {I^{+}\left({k}\right)} \sum\limits_{l=0}^{4}{I^{-}\left({l}\right)} }\right] - 2 \mathbf{1}\} = 
$$
\nopagebreak \begin{equation}
\label{eq71}
 = - \left({\omega_{I}-a/2}\right)\sum\limits_{k=0}^{4} {I_{z}\left({k}\right)} + U\{            \sum\limits_{k=0}^{4} {I    \left({k}\right)} \sum\limits_{l=0}^{4}{I    \left({l}\right)} - \sum\limits_{k=0}^{4} {I_{z}\left({k}\right)} \sum\limits_{l=0}^{4}{I_{z}\left({l}\right)}          - 2 \mathbf{1}\}
\end{equation}

The identity component of $h_{xxo}^{4}$ in Eq.(\ref{eq71}) is not relevant here and will be next omitted.

The eigenvalues of such Hamiltonian are 
\begin{equation}
\label{eq72}
E\left({J,m_{J}}\right) = - |\omega_{I} - a/2|m_{J} + U\left({J\left({J + 1}\right) - m_{J} ^{2}}\right),\,\,\,m_{J} = 0,\,\, \pm 1,\,\,\,.... \pm J,\,\,\,J = 0,\,\,1,\,\,2.
\end{equation}

They are tabulated below in Table:
\begin{table}[ht]
  \centering
   \begin{tabular}[c]{c|c|c|c}
                    & $ J=0$ &          $J=1$                &              $J=2$            \\ \hline
     $m_{J}=0\,\,\,$  &  $0$ &          $2U$                 &               $6U$            \\
     $m_{J}=\pm 1  $  &      &   $\pm|\omega_{I} - a/2| + U$ & $\pm|\omega_{I} - a/2| + 5U$  \\
     $m_{J}=\pm 2  $  &      &                               &2$\pm|\omega_{I} - a/2| + 2U$ 
   \end{tabular}
  % \caption{caption}
  \label{tab:tab1}
\end{table}

Because that $|\omega_{I} - a/2| \gg U\sim a^{2}$, 
the states of four considered physical qubits here represent approximate so-named weak 
collective decoherence free subspace (Ref.[9]). 
The ground state of Hamiltonian (\ref{eq71}) corresponds to nondegenerate state with 
$J = 2$, $m_{J} = - 2$ and to eigenvalue $ - 2 |\omega_{I} - a/2| + 2U$. 
It can not use for two logical qubit encoding.

However the all states with $m_{J} = 0$ 
are identical to the strong collective DFS state with 
$J = 0$, $m_{J} = 0$ 
which overlie the ground state by $2\,|\omega_{I} - a/2| - 2U$. 
It is represented by two-fold degenerated state of Eq.(\ref{eq67}) type. 
This state at low temperatures 
$|\omega_{I} - a/2| \gg T/\left({\hbar \omega_{E} /k_{B}}\right)$ ($T\sim 1\,\mathrm{mK}$)
is a metastable state and, consequently, can be used as two DFS encoded logical qubits.

\section*{Conclusion}

It was considered the results of theoretical investigations of one qubit 
and two qubit nonadiabatic decoherence and longitudinal relaxation caused by interaction 
of nuclear spins-qubits with virtual magnon excitations in antiferromagnet. 
It turns out that the character of decoherence processes essentially depends on 
antiferromagnet anisotropy (parameter $b_{C} $) and on inhomogeneity of external field (parameter $g$). 
As this takes place, the temperature whereby the thermal magnon excitations are excluded 
and the two-magnon spin-lattice relaxation is especially suppressed, 
should be defined by values 
$T \ll T_{C} \left({1 - b/b_{C}}\right)$, 
$T_{C} = \hbar \gamma_{S} B_{C} /k_{B}$. 
As an example we have also considered decoherenc of pair qubits maximally entanglement state 
and have calculated the concurrence damping rate.

The other mechanism of quantum state decoherence is adiabatic process 
of the resonance nuclear spin frequency modulation caused by dipole-dipole interaction 
with nuclear spins of impurity isotopes. 
The necessary degree of the adiabatic decoherence suppression can be obtained 
at spin temperature less than 1mK and for the concentration of impurity nuclear spin 
containing isotopes less than $10^{-2}\%$.

Finally it was discussed for considered model of quantum register 
the possibility of construction of DFS-states.

\end{document}